\documentclass[twocolumn,prb,showpacs]{revtex4}
\usepackage{graphicx}
\usepackage{amsmath,amssymb,amsfonts}
\usepackage{bm}
\usepackage{hyperref}

\begin{document}

\title{Orbital-Ordering Induces Structural Phase Transition and the Resistivity Anomaly in Iron Pnictides}

\author{Weicheng Lv, Jiansheng Wu, and Philip Phillips}
\affiliation{Department of Physics, University of Illinois at Urbana-Champaign, 1110 West Green Street, Urbana, IL 61801}
\date{\today}

\begin{abstract}
We attribute the structural phase transition (SPT) in the parent compounds of the iron pnictides to orbital ordering. Due to the anisotropy of the $d_{xz}$ and $d_{yz}$ orbitals in the $xy$ plane, a ferro-orbital ordering makes the orthorhombic structure more energetically favorable, thus inducing the SPT. In this orbital-ordered system, the sites with orbitals that do not order have higher energies. Scattering of the itinerant electrons by these localized two-level systems causes a resistivity anomaly upon the onset of the SPT. The proposed orbital ordering also leads to the stripe-like anti-ferromagnetism and anisotropy of the magnetic exchanges. This model is quantitatively consistent with available experimental observations.
\end{abstract}


\maketitle

\section{Introduction}

The structural phase transition (SPT) from tetragonal to orthorhombic
symmetry around $150\,\mathrm{K}$ \cite{Nomura} is a ubiquitous feature in
the parent compounds of the iron-based superconductors. Coincident
with this transition is a resistivity anomaly (RA) \cite{Dong_epl} in
which the resistivity turns up slightly before a sharp drop at exactly
the onset temperature of the SPT, $T_\mathrm{SPT}$. For the
1111-family, at a lower temperature, $T_\mathrm{SDW}$, a stripe-like
anti-ferromagnetic spin density wave (SDW) forms \cite{Cruz_nature_sdw}
on the distorted lattice of Fe atoms, with the spins being parallel
along the shorter axis and anti-parallel along the longer
axis. However, for the 122-family, the SDW develops at the same
temperature as does the SPT, $T_\mathrm{SDW} = T_\mathrm{SPT}$
\cite{Huang_prl_122}. In the 122-family \cite{Goldman_122_pt}, a
single first-order transition obtains instead of two separate
second-order transitions in the 1111-family. Upon doping,
superconductivity (SC) occurs leading to a cessation of the SPT, RA
and SDW \cite{Zhao_Ce_pd, Luetkens_La_pd}. Hence, all of these three phenomena should be closely related and share a universal mechanism. However, most theoretical work only focuses on the connections between the SDW and SC. The importance of the SPT and RA is somehow underestimated. The main objective of this paper is to explain the origin of the SPT and RA.

A common view \cite{SpinAngle, SpinLength,Xu,Hu} is that the SPT is driven by the onset of the stripe-like anti-ferromagnetism.  Both first principles calculations \cite{SpinAngle,SpinLength} and Landau-Ginzburg modelings \cite{Xu,Hu} have been used in this context. The fact that the two transitions are decoupled
in the 1111-family is a limitation of this approach.  Further, since the origin of the SPT in their scenario is spin based, the onset temperature should be sensitive to an external
magnetic field. However, experiments have shown that varying the
magnetic field leads to no change in the onset temperature of the
SPT \cite{Dong_epl}. 

In this paper, we develop a microscopic
theory of the SPT without involving the spin degrees of freedom. 
On our account, uneven occupations of the $d_{xz}$ and $d_{yz}$ orbitals
make the orthorhombic crystal structure more energetically favorable, thus inducing the SPT. 
The operative mechanism driving this ferro-orbital-ordering transition is the lifting of the degeneracy between the $d_{xz}$ and $d_{yz}$ orbitals by the inter-site Coulomb repulsions. However, it should be noted that other important factors, such as spin-orbit interactions \cite{KK} and couplings to the displacements of ligand atoms (As), also contribute to this process. In fact, spin-orbit physics appears to lie at the heart of orbital ordering in the manganites\cite{Tokura}. While such physics is undoubtedly present in the pnictides \cite{Kruger_orb,Singh}, quantifying it would require a first-principles calculation of the relevant parameters.  However, as our goal is to propose a simple mechanism that explains both the SPT and the resistivity anomaly, we focus on a more easily quantifiable approach to orbital ordering based instead on the Coulomb repulsion. Indeed, what our work indicates is that there is a rich set of models which can lead to orbital ordering in the pnictides. Our model is sufficiently simple and general that warrant its being taken seriously. The key insight gained from this study is not the detailed microscopic mechanism for this orbital-ordering-induced SPT, which is rather standard \cite{Tokura}, but its direct consequence - a resistivity anomaly, which can be captured by our model in quantitative agreement with the experimental results [see Fig.~\ref{res}(b)]. Furthermore, the stripe-like SDW and recently discovered anisotropy \cite{Zhao_aniso} of the magnetic exchanges naturally arise in our theory.

\section{Orbital Ordering}
\label{sec_spt}
\begin{figure}
  \centering
  \includegraphics[width=8cm]{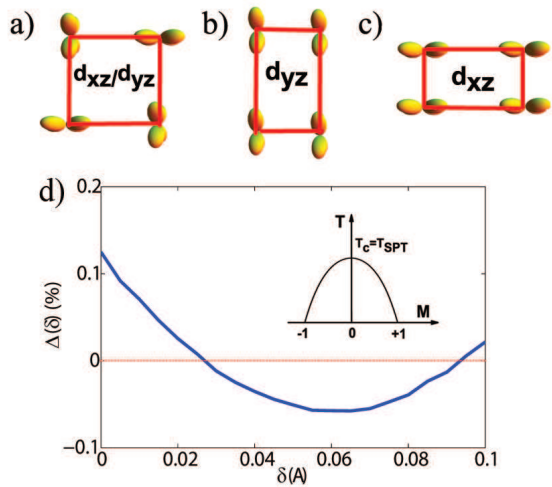}
  \caption{(Color online) (a) Equal numbers of $d_{xz}$ and $d_{yz}$ with a square lattice configuration. (b) Entirely $d_{yz}$ state with $a<b$. (c) Entirely $d_{xz}$ state with $a>b$. d) Relative energy difference $\Delta$ between configurations (a) and (b), or (c), as a function of lattice distortion $\delta$. (Inset: an Ising-type transition where the order parameter $M$ is defined as the difference between the numbers of occupied $d_{xz}$ and $d_{yz}$ orbitals.) }
  \label{ising}
\end{figure}

As being emphasized by pioneering earlier work \cite{Kruger_orb}, the orbital degrees of
freedom are important in the iron pnictides, which are intrinsically
multi-orbital systems. For the Fe atom located at the center of the
tetrahedron of four neighboring As atoms, the five $d$ orbitals are
split into two groups, $t_{2g}$ ($d_{xy}$, $d_{xz}$, $d_{yz}$) and
$e_g$ ($d_{x^2-y^2}$, $d_{3z^2-r^2}$). Three of the five orbitals,
$d_{xy}$, $d_{x^2-y^2}$ and $d_{3z^2-r^2}$ are rotationally symmetric in
the $xy$ plane. So they are unlikely to have any effect on the SPT
which is asymmetric in the $xy$ plane. Then the only two possible
candidates are the $d_{xz}$ and $d_{yz}$ orbitals.  We propose the
following  mechanism for the SPT, assuming these two orbitals are
localized. At high temperature $T>T_{\mathrm{SPT}}$, $d_{xz}$ and
$d_{yz}$ orbitals are degenerate, with equal numbers of electrons on
both.  A possible configuration is shown in Fig.~\ref{ising}(a), in
which a square lattice is preferred. At low temperature,
$T<T_{\mathrm{\rm SPT}}$, there is a majority of either $d_{xz}$ or
$d_{yz}$. For $d_{yz}$ orbitals, the 
Coulomb repulsion of two neighboring sites is stronger along the
$y$-direction than along the $x$-direction, which leads to a
rectangular lattice with $a<b$ as shown in Fig.~\ref{ising}(b), where
$a$ and $b$ are unit lengths in the $x$ and $y$ direction,
respectively. Similarly, when $d_{xz}$ dominates, the system will form the configuration of Fig.~\ref{ising}(c), which is degenerate with (b) by a rotation of 90 degrees.

To demonstrate the viability of this mechanism, we need to compare the energies of configuration (a) and (b) in Fig.~\ref{ising}. 
For simplicity, only the nearest-neighbor Coulomb repulsions are considered,
\begin{equation}
	U = \int d\bm{r} \, d\bm{r^{\prime}} \frac{e^2}{\left| \bm{r} - \bm{r^\prime}\right|}
	\left| \psi_{\alpha}(\bm{r} - \bm{R}_i) \right| ^2
	\left| \psi_{\beta}(\bm{r^\prime} - \bm{R}_j) \right| ^2
\end{equation}
where $\psi_{\alpha}(\bm{r} - \bm{R}_i)$ is the wave function of the $\alpha$ ($\alpha = d_{xz}, d_{yz}$) orbital electron at site $\bm{R}_i$. This integral can be evaluated by an importance-sampling Monte-Carlo method.
In configuration (a), we choose $a=b=a_0=2.85$ \AA, which is the typical experimental value \cite{Nomura} for the 1111-family. For configuration (b), we define the lattice distortion $\delta$ as $a=a_0-\delta$, and set $b=a_0^2/a$ to keep the area of a unit cell constant. We calculate the relative energy difference
\begin{equation}
	\Delta (\delta) = \frac{U_b (\delta) - U_a}{U_a}
\end{equation}
as a function of $\delta$, where $U_a$ and $U_b$ are energies of
configurations (a) and (b) respectively. The results are shown in
Fig.~\ref{ising}(d). For a lattice distortion $0.03$ \AA $<\delta<0.09$ \AA, the rectangular lattice (b) or (c) is more energetically favorable. It is noted that this value is larger than the experimentally observed distortion of about 0.01 \AA \cite{Nomura}. However, the localized states are probably neither $d_{xz}$ nor $d_{yz}$, but some combinations of the $d$ orbitals, or even involve hybridization with As $p$ orbitals \cite{Wu_moment}. Thus the precise value of the distortion length can be smaller by taking these factors into account. As already mentioned, other possibilities may also induce this ferro-orbital ordering and the subsequent SPT. For example, Kr\"uger {\it et. al} \cite{Kruger_orb} derive a Kugel-Khomskii spin-orbital model and the resultant phase diagram does contain the same orbital configuration as proposed here. However, constructing the complete microscopic Hamiltonian that incorporates all the important physical processes requires a detailed knowledge of the relevant coupling parameters, which is currently unavailable.  Thus the key point of our study is to put forth a simplified picture based on the coupling only to the Coulomb interaction in which the rectangular lattice with ferro-orbital ordering emerges spontaneously at low temperature because of its lower energy. 

Our model allows us to make the following conclusion. Upon the onset of the phase transition, a lattice distortion breaks the degeneracy between $d_{xz}$ and $d_{yz}$. By occupying either one of these two orbitals, the system forms a ferro-orbital-ordered state and lowers its energy. It is this orbital-driven Jahn-Teller effect that induces the SPT. Defining $M_i = \pm 1$ for site $i$ occupied by $d_{xz}$ and $d_{yz}$ orbitals respectively, we can write down an effective Ising-type Hamiltonian for the SPT
\begin{equation}
H_\mathrm{SPT} = - J_\mathrm{SPT} \sum_{\langle i,j\rangle} M_i M_j
\end{equation}
where $J_\mathrm{SPT}$ should be on the order of the transition temperature, $T_{\mathrm{SPT}}$. So the SPT belongs to the Ising universality class, as shown in the inset of Fig.~\ref{ising}(d), where the order parameter $M$ is defined as $M = \sum_{i} M_i / N$.

Recently, angle-resolved photoemission experiments using a linear-polarized laser beam \cite{ARPES_orb} show that at low temperature, the Fermi surface at the Brillouin zone center is dominated by a single $d_{xz}$ or $d_{yz}$ orbital, depending on the distortions. In their subsequent local-density approximation (LDA) calculations \cite{ARPES_orb}, it is found that the density of states of the $d_{yz}$ orbitals with a lattice configuration of $a<b$ displays a peak around $0.5\,e\rm{V}$ from the chemical potential, which is just the localized state predicted in our SPT model. A recent optical measurement \cite{optical} also suggests evidences of the orbital ordering.  

\section{Resistivity Anomaly}

The ferro-orbital-ordering-driven SPT mechanism has an important consequence, namely the resistivity anomaly.  The essential physics is that of a Kondo problem. The scattering of the itinerant electrons off two otherwise degenerate orbitals, $d_{xz}$ and $d_{yz}$, will be suppressed by the gap opening, which results in a sharp drop of the resistivity upon the onset of the SPT. 

\begin{figure}
  \centering
  \includegraphics[width = 8.0cm]{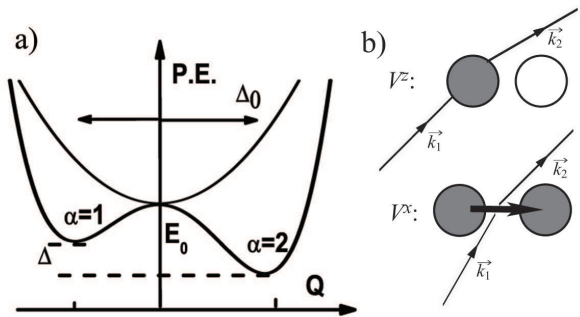}
  \caption{(a) A schematic of a double-well potential, as the classical analog of the two-level system. (b) Two types of scattering processes between the itinerant electrons and the localized states. $V^z$: diagonal scattering; and $V^x$: off-diagonal scattering.}
  \label{tls}
\end{figure}

Above, $T_{\rm SPT}$, the two $d_{xz}$ and $d_{yz}$ orbitals are degenerate. Below $T_{\rm SPT}$, the occupancy of the electrons in  $d_{xz}$ and $d_{yz}$ orbitals becomes unbalanced as a result of the distortion of the crystal
to configuration (c) [or (b)] in Fig.~\ref{ising}. Thus, the electrons
that remain in the $d_{yz}$ (or $d_{xz}$) orbitals will have a higher
energy and hence can lower their energy by jumping onto $d_{xz}$
(or $d_{yz}$) orbitals.  This process can be described by a localized
two-level system.  The classical analog, namely a double-well
potential, is shown in Fig.~\ref{tls}(a).  The corresponding Hamiltonian is given by 
\begin{eqnarray}
	 H_{\rm TLS} &=& \lambda_{\mathrm{ps}} \sum_{\alpha} a_{\alpha}^{\dag} a_{\alpha} + \frac{1}{2} \Delta \sum_{\alpha \beta} a_\alpha^\dag \sigma_{\alpha\beta}^z  a_\beta  \nonumber \\
  & & + \frac{1}{2} \Delta_0 \sum_{\alpha \beta} a_\alpha^\dag \sigma_{\alpha\beta}^x a_\beta 
 \label{eq:tls}
\end{eqnarray}
where $a_{\alpha}^{\dagger}$ ($a_{\alpha}$) creates (annihilates) an electron on orbital $\alpha$ and $\sigma^i_{\alpha \beta}$ is a Pauli matrix. We will choose an appropriate fictitious energy $\lambda_{\mathrm{ps}}$ to prevent the system from double occupancy. $\Delta$ is the energy splitting between the two levels and $\Delta_0$ is the tunneling rate, as shown in Fig.~\ref{tls}(a). By a rotation of the spin axis, this system can be diagonalized and the gap between the two eigenstates is
$E = \sqrt{ \Delta_0^2 + \Delta^2 }$.

As the parent compounds are actually metallic, there should be
itinerant electrons present besides these localized states. These two
can be coupled as in the framework of the localized-itinerant
dichotomous models\cite{Wu_moment,Si_coh-inc,Weng,Wu_magnon}. The starting Hamiltonian is \cite{2level}
\begin{eqnarray}
  H &=& H_e + H_{\rm TLS} + V \label{eq:all} \\
  H_e &=& \sum_{k \sigma} E_k c_{k \sigma}^{\dag} c_{k \sigma} \label{eq:itin} \\
  V &=& \sum_i \sum_{k_1 \sigma_1, k_2 \sigma_2} \sum_{\alpha \beta}  c_{k_2 \sigma_2}^{\dag} V_{k_2 k_1}^{i} c_{k_1 \sigma_1} \ a_\alpha^\dag \sigma_{\alpha\beta}^i  a_\beta
\end{eqnarray}
where $H_e$, $H_{\rm TLS}$ and $V$ represent the Hamiltonians for the
itinerant electrons, the single two-level system and the interactions
between the two, respectively. There are two kinds of scattering
processes as shown in Fig.~\ref{tls}(b). One is the diagonal
scattering described by the $V^z$ term, where the localized state
remains on the same level. The other is the off-diagonal scattering
initiated by the $V^x$ term, where the localized state jumps onto the other level. $V^y$ is in fact zero, as it breaks time-reversal symmetry. However, it should be noted that $V^y = 0$ does not hold for the renormalized vertex, since higher order terms are not necessarily local. We will also assume $V^z \gg V^x$ as proposed previously \cite{2level}.

In fact, this system is very similar to the Kondo model, with the two
orbitals $d_{xz}$ and $d_{yz}$ representing the up and down-spin states on the magnetic impurity. We are going to perform a similar
scaling analysis following Ref.~\onlinecite{2level}. We define the
dimensionless couplings $v_{k_1 k_2}^i = V_{k_1 k_2}^i N_0$ where
$N_0$ is the density of states at the Fermi level. Reducing the
bandwidth from $D_0$ to $D$ and evaluating the vertex corrections up to the leading order, we have the scaling equations
\begin{eqnarray}
	\frac{\partial v_{\alpha \beta}^s(u)} {\partial u} & = & - 2i \sum_{ij} \sum_{\gamma} \epsilon^{ijs} v_{\alpha \gamma}^i(u) v_{\gamma \beta}^j (u) 
\end{eqnarray}
where $v_{\alpha \beta} ^i$ are defined as $v_{k_1 k_2}^i = \sum
f_\alpha^\dag (\hat{k_1}) v_{\alpha \beta} f_\beta (\hat{k_2})$, with
$f_\alpha (\hat{k})$ being a 
complete set of spherical harmonics, $f_\alpha (\hat{k}) = i^l Y_l^m (\theta_k, \phi_k)$, $\epsilon^{ijs}$ is the Levi-Civita symbol and $u = \ln (D/D_0)$. We can express $v_{\alpha \beta}^i$ using the Pauli matrices as $v_{\alpha \beta}^i = v^i \sigma_{\alpha \beta}^i$. Then the above scaling equations will be reduced to a set of coupled equations involving $v^x$, $v^y$ and $v^z$. These equations can be solved by separating $u$ into two regimes: (a) $v^y < v^x \ll v^z$ and (b) $v^y \simeq v^x < v^z$. In regime (a), the solutions are
\begin{eqnarray}
	v^x (u) & = & v^x(0) \cosh \left[ 4v^z(0) u\right] \\
	v^y (u) & = & v^x(0) \sinh \left[ 4v^z(0) u\right] \\
	v^z (u) & = & v^z(0).
\end{eqnarray}
In regime (b), we have
\begin{eqnarray}
	 \left[ v^z(u) \right] ^2 - \left[ v^x(u) \right] ^2 = v_0^2,
\end{eqnarray}
where $v_0$ is scale invariant and $v^z(u)$ satisfies
\begin{eqnarray}
	u = - \frac{1}{4v^z(u)} + \ln \left[ \frac{D_0}{k_\mathrm{B} T_k}\right] 
\end{eqnarray}
with the Kondo temperature $T_k$ identified as
\begin{eqnarray}
	k_\mathrm{B} T_k = D_0 \left[ \frac{v^x(0)}{4v^z(0)} \right] ^{1/4v^z(0)}.
\end{eqnarray}

\begin{figure}
  \centering
  \includegraphics[width = 8.0cm]{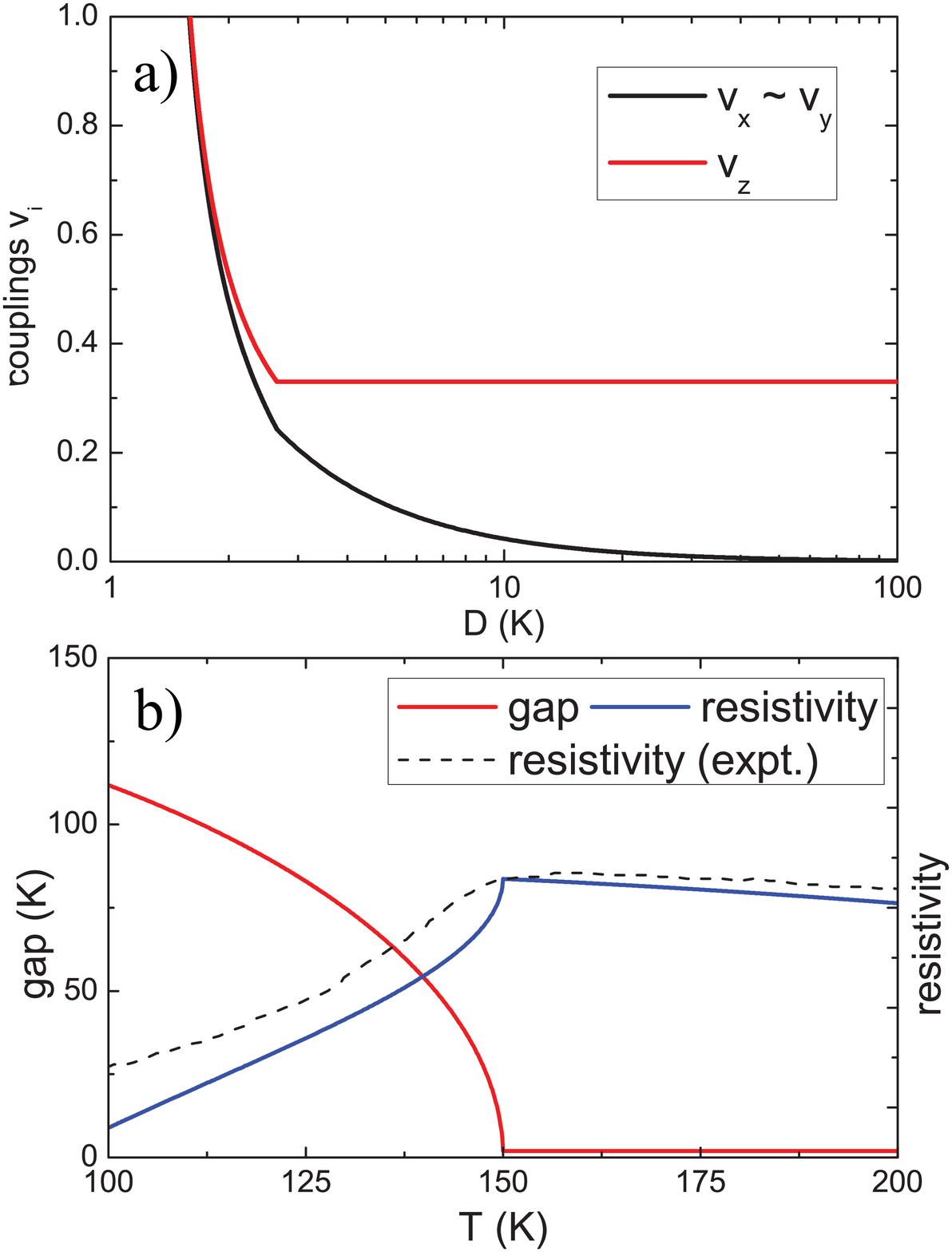}
  \caption{(Color online) (a) Scaling of the coupling constants $v^i$ with respect to bandwidth $D$. (b) Energy gap and resistivity as a function of temperature $T$. (The experimental data of resistivity are extracted from Ref.~\cite{Dong_epl}.) Setting the resistivity at $T=150K$ of our model equal to that of the experiment was the only fitting parameter.}
  \label{res}
\end{figure}

Using the parameters $v^z(0) = 0.33$, $v^x(0)/v^z(0) = 0.001$, $v^y(0) =0$ and $D_0 = 665 \, \mathrm{K}$ \cite{Kata}, we obtained the scaling flows of $v^x$, $v^y$ and $v^z$ shown in Fig.~\ref{res}(a), for $E = 0 \, \mathrm{K}$. The corresponding Kondo temperature is $T_k = 1.24 \, \mathrm{K}$. Reducing the bandwidth $D$, the system goes from weak to strong coupling. The resistivity due to the scattering of the two-level system can be calculated based on these renormalized vertices as in Ref.~\onlinecite{Kata}. 
At high temperature, we have two degenerate levels, $d_{xz}$ and
$d_{yz}$. When the temperature is reduced, the scattering from the states closer
to the chemical potential increases, leading to a resistivity upturn of $\log{T}$
\cite{kondo_behavior} as in the Kondo model. However, upon the onset
of the SPT, a gap opens between the two levels. If the bandwidth $D$
is less than the gap $E$, the off-diagonal scattering is not allowed,
since there are no states for the electrons to be scattered into. As a
consequence, the scaling terminates at $D=E$. The
electrons within the bandwidth $E$ will no longer contribute to the resistivity. This is the mechanism behind the resistivity anomaly. Our result is shown in Fig.~\ref{res}(b), which is in good qualitative agreement with experiment. We set the tunneling rate $\Delta_0 = 2 \, \mathrm{K}$, and the energy splitting takes the form 
\begin{equation}
\Delta(T) = \Delta(0)  \sqrt{1 - \left( \frac{T}{T_\mathrm{SPT}} \right)^2}
\end{equation} 
where $\Delta(0) = T_\mathrm{SPT} = 150 \, \mathrm{K}$
when $T< T_\mathrm{SPT}$. 
It should be noted that the overall behavior of the scaling flows and the resistivity are independent of the chosen parameters. This represents the explanation of the RA for the iron pnictides.

\section{Orbital Driven Magnetism}

\begin{figure}
  \centering
  \includegraphics[width=8cm]{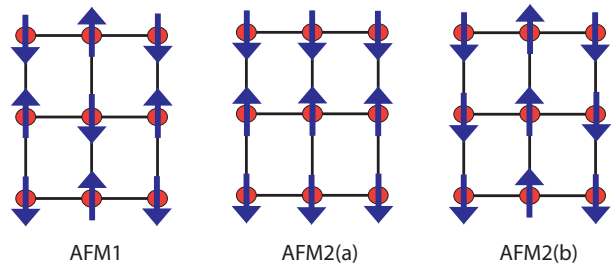}
  \caption{(Color online) Different possible spin configurations on a distorted lattice with $a<b$, which corresponds to the case that $d_{yz}$ is the majority orbital.}
  \label{afm}
\end{figure}

Our model also offers a natural solution to the observed stripe-like anti-ferromagnetism.
Before the SPT, we have an orbitally disordered state, in
which the neighboring sites are occupied probabalistically by
different orbitals.  The resultant lack of overlap gives rise to a vanishing of
any anti-ferromagnetic spin exchange and as a consequence no spin
order. After the SPT, either $d_{xz}$ or $d_{yz}$ orbitals will
dominate. Without loss of generality, we suppose that most sites
are occupied by $d_{yz}$, as shown in Fig.~\ref{ising}(b). Due to the
larger overlap of the wave functions on neighboring sites in the
$y$-direction than that in the $x$-direction, the hopping integral
$t_b$ should be larger than $t_a$. For the nearest-neighbor spin
exchange, $J_1 \sim t^2/U$, we have that $J_{1a} < J_{1b}$. So the spins on the longer axis have a stronger tendency to be aligned oppositely. The spin configuration AFM2(b) in Fig.~\ref{afm} is not favored. As has been suggested\cite{SpinAngle,j1j2}, we can further introduce a next-nearest-neighbor exchange $J_2$. If $J_2 > J_{1a}/2$, which is very likely for a relatively small $J_{1a}$ \cite{Kruger_orb}, AFM2(a) will have a lower energy than AFM1, as shown in Fig.~\ref{afm}, and emerge as the ground state at low temperature, which has already been confirmed by the experiments\cite{Cruz_nature_sdw}. In contrast with other theories in which the SPT is induced by the spin degrees of freedoms, on this account, the formation of the SDW is actually a result of the ferro-orbital ordering accompanying the SPT. 

In fact, we are able to construct a universal Hamiltonian describing both the SPT and SDW, with a spin-orbit coupling model \cite{Tokura,Kruger_orb, Singh},
\begin{eqnarray}
	H_\mathrm{SO} & = & J_{\rm{SPT}} \sum_{\langle i,j \rangle} M_i M_j + \sum_{\langle \langle i,j \rangle \rangle} J_{2}\left(M_i,M_j \right) \bm{S}_i \cdot \bm{S}_{j}  \nonumber \\
	 && + \sum_i J_{1x}\left(M_i,M_{i+\hat{x}} \right) \bm{S}_i \cdot \bm{S}_{i+\hat{x}} \nonumber  \\
	 && +  \sum_i J_{1y}\left(M_i,M_{i+\hat{y}} \right) \bm{S}_i \cdot \bm{S}_{i+\hat{y}} 
\end{eqnarray}
where the spin exchanges are given by
\begin{eqnarray}
	J_{1x}\left(M_i,M_j \right) & = & \delta_{M_i,M_j} \left( J_{1b} \delta_{M_i,1} + J_{1a} \delta_{M_i, -1} \right) \\
	J_{1y}\left(M_i,M_j \right) & = & \delta_{M_i,M_j} \left( J_{1a} \delta_{M_i,1} + J_{1b} \delta_{M_i, -1} \right) \\
	J_2\left(M_i,M_j \right) & = & \delta_{M_i,M_j} J_2
\end{eqnarray}
where $M_i$, representing the orbital degrees of freedom, is defined to be $\pm 1$ for $d_{xz}$ and $d_{yz}$ respectively,  as in Sec.~\ref{sec_spt}.
Clearly, in this model, the spin order will not occur until the formation of the ferro-orbital ordering at $T_\mathrm{SPT}$, which is on the order of $J_\mathrm{SPT}$. Below $T_\mathrm{SPT}$, the spin degrees of freedom can be described by an anisotropic Heisenberg model, whose transition temperature to the spin-ordered state, $T_s$, would depend on the spin exchanges, $J_{1a}$, $J_{1b}$ and $J_2$. If $T_S < T_{\mathrm{SPT}}$, we would
have two separate second-order transitions, $T_{\mathrm{SDW}} = T_S
<T_{\mathrm{SPT}}$, as in the case of the 1111-family. For the
122-family, which has a shorter Fe-Fe bond length, it is expected this would enhance
the spin exchange $J$, likely leading to $T_S > T_{\mathrm{SPT}}$. But the SDW will not form before the SPT, since there is no spin exchange until the SPT obtains. So there is only one first-order transition, $T_{\mathrm{SDW}} = T_{\mathrm{SPT}}$.  

Furthermore, this anisotropic Heisenberg model has also been proposed on experimental grounds~\cite{Zhao_aniso} to fit the spin-wave spectrum seen in the inelastic neutron scattering
data. Our theory gives a direct explanation for the observed anisotropy of magnetic exchanges. Note their results~\cite{Zhao_aniso} do rely on a negative $J_1^a$, which is not
obtained by our simple model. However, this difficulty can overcome by introducing a Hund's coupling between these localized spins and itinerant electrons,
\begin{equation}
	H_K = -\frac{J_H}{2} \sum_{i,\nu \nu^\prime} \bm{S}_i \cdot c_{i\nu}^\dagger \bm{\sigma}_{\nu \nu^\prime} c_{i\nu^\prime}
\end{equation} 
where $\bm{\sigma}_{\nu \nu^\prime}$ are the Pauli matrices.  The hopping of the itinerant electrons with this Hund's coupling will give rise to an effective ferromagnetic coupling \cite{Zener, Anderson, Gennes} between neighboring spins. After taking this into account, we will eventually have the spins on the shorter axis coupled ferromagnetically. The full details of this model are the subject of a future study.

\section{Final Remarks}

To conclude, we have proposed that the SPT and RA in the iron pnictides are due to the opening of a gap between two otherwise degenerate orbitals. While our mechanism for the structural phase transition is a standard Jahn-Teller distortion driven by a minimization of the Coulomb repulsion, the key point of this paper is that the resulting simple two-level system can resolve the previously unexplained resistivity anomaly. The mechanism proposed here is
independent of an applied magnetic field as is seen experimentally\cite{Dong_epl}.  Only
in a ferro-orbital-ordered state after the SPT does the stripe-like SDW form. This is the reason why these three phenomena, SPT, RA and SDW, are closely related and almost always coincide with one another. In doped materials, extra electrons or
holes will break the uneven occupations of $d_{xz}$ and $d_{yz}$, thus
diminishing the Jahn-Teller effect. So the SPT, RA and SDW will all
become less pronounced and shift to lower temperature, eventually
vanishing at some critical doping. These are all observed
experimentally, lending credence to our model. 

After this work was posted on arXiv, several similar papers\cite{Ku, Turner} appeared, based on the same orbital physics we utilized here, which supports our theory that the orbital ordering is the driving mechanism for the SPT, RA and SDW. 

\begin{acknowledgments}
We thank the NSF under Grant No.~DMR-0605769 for partial funding of this work and Frank Kr\"uger for his insightful remarks.
\end{acknowledgments}

\end{document}